\documentclass{elsarticle}
\usepackage{amsmath}
\usepackage{ecrc}
\usepackage[bookmarks=false]{hyperref}
    \hypersetup{colorlinks,
      linkcolor=blue,
      citecolor=blue,
      urlcolor=blue}

\firstpage{1}

\usepackage{amssymb}

\usepackage{multicol}
\usepackage{tikz}
\usetikzlibrary{positioning}
\usepackage{booktabs}

\usepackage[figuresright]{rotating}

\begin{document}
\begin{frontmatter}




\title{Forecasting Empty Container availability for Vehicle Booking System Application}


\author[a]{Arthur Cartel Foahom Gouabou\corref{cor1}} 
\author[a]{Mohammed Al-Kharaz\corref{cor1}}
\author[a,b]{Faouzi Hakimi} 
\author[a,b]{Tarek Khaled}
\author[a]{Kenza Amzil}

\address[a]{DMSLOG AI, 5971 W 3rd St, Los Angeles, CA, USA}
\address[b]{Aix-Marseille University, 58 Blvd Charles Livon, 13007 Marseille, France}

\begin{abstract}

Container terminals, pivotal nodes in the network of empty container movement, hold significant potential for enhancing operational efficiency within terminal depots through effective collaboration between transporters and terminal operators. This collaboration is crucial for achieving optimization, leading to streamlined operations and reduced congestion, thereby benefiting both parties. Consequently, there is a pressing need to develop the most suitable forecasting approaches to address this challenge. This study focuses on developing and evaluating a data-driven approach for forecasting empty container availability at container terminal depots within a Vehicle Booking System (VBS) framework. It addresses the gap in research concerning optimizing empty container dwell time and aims to enhance operational efficiencies in container terminal operations. Four forecasting models—Naive, ARIMA, Prophet, and LSTM—are comprehensively analyzed for their predictive capabilities, with LSTM emerging as the top performer due to its ability to capture complex time series patterns. The research underscores the significance of selecting appropriate forecasting techniques tailored to the specific requirements of container terminal operations, contributing to improved operational planning and management in maritime logistics.
\end{abstract}

\cortext[cor1]{Corresponding author}
\end{frontmatter}




\section{Introduction}
\label{introduction}
Maritime transport is the backbone of global trade, facilitating the worldwide shipping of goods across oceans. The
rapid growth of containerised trade on a global scale, coupled with the emergence of mega vessels, has led
to a significant annual increase in container throughput. Port's Container Terminals function as pivotal hubs within the global supply chain, providing temporary storage for containerized cargo in terminal yards before onward distribution. Given their crucial role as key nodes in the global container distribution network, the reliability and efficiency of these terminals significantly influence the prosperity and advancement of the global supply chain. The empty container assumes a pivotal role in optimizing port operations within the broader supply chain framework. Its indispensability lies in its ability to facilitate port activity and the uninterrupted movement of goods around the world. Over time, the management of empty containers has emerged as a specialized domain within the supply chain ecosystem since the inception of global containerization \cite{zain2014understanding}. The primary activity involving empty containers is the physical movement of items and products, orchestrated by logistics providers including shipping lines, transporters, freight forwarders, and other entities offering logistic services \cite{mol2007logistics}. Notably, key stakeholders in the logistics services landscape can be identified based on their specific roles, encompassing importers, exporters, road carriers/hauliers, shipping lines/sea carriers, empty container depots, forwarders, customs and regulatory authorities, and inland terminal operators. As such, these entities collectively play a crucial role in orchestrating the logistics and management of empty container movements \cite{karmelic2012empty}.  \\

The literature on empty container logistics primarily focuses on addressing Empty Container Repositioning (ECR) challenges, covering global, inter-regional, and regional planning strategies \cite{Finke2016, genccer2019overview, hanafi2023mathematical, boile2006empty, mittal2008regional}. For a comprehensive exploration of ECR research and trends, interested readers are referred to \cite{genccer2019overview, abdelshafie2022repositioning}. Recent advances in
in data-driven approaches and machine learning offer liner shipping companies the ability to optimise container
container movements, anticipate surpluses or deficits and implement dynamic pricing strategies \cite{martius2022forecasting}. This proactive approach facilitated by predictive modeling minimizes costs in transportation operations. In addition, the integration of Artificial Intelligence (AI) techniques  into port planning and operations extends to the forecasting of port container throughput \cite{yang2020forecasting, awah2021short, munim2023forecasting}, container shipping demand \cite{ubaid2021container}, and freight rates \cite{saeed2023forecasting}, respectively, to support informed decisions
for infrastructure expansion, business decisions, sales optimisation and price management. Notably, attention is focused on enhancing decisions concerning empty container relocation through accurate forecasting of empty container availability \cite{diaz2011forecasting, liu2019machine, meseguer2021artificial, martius2022forecasting, cai2022data, sommer2023forecasting}, crucial for proactive repositioning decisions and cost reduction within port areas. Various methods have been developed and implemented to forecast empty container availability, aimed at optimizing repositioning decisions. However, until recently, there has been a gap in research concerning the optimization of empty container dwell time within terminal depots. Container terminals, as key nodes in the in the empty container movement network, have the potential to improve empty container flow management by minimising dwell time and reactive movements within the terminal depot. Achieving this optimisation requires collaboration and information sharing between carriers and terminal operators. Transporters benefit from streamline operations and maximise efficiency during terminal visits, while terminals improve operational efficiency and reduce congestion. It is assumed that container terminals use a Vehicle Booking System (VBS) to facilitate the organisation and planning of collection and delivery operations for full and empty containers. Modern VBSs are increasingly integrating data-driven methods into their truck arrival management systems, highlighting the need for AI techniques such as forecasting methods. The objective of this paper is to develop a data-driven approach for forecasting the availability of empty containers at container terminal depots per VBS time slot for a future time horizon of five business days.\\  

The remainder of this paper is organized as follows: Section 2 provides a brief overview of the existing literature on forecasting methodologies in empty container management. Section 3 explains the proposed approach for forecasting empty container availability within container terminals. In Section 4, we present and discuss the results, demonstrating the effectiveness and applicability of the proposed approach through a comprehensive evaluation using well-defined metrics. Finally, in Section 5, we conclude the study with an outline of possible avenues for future research in the area of empty container availability forecasting.

\section{Literature Review}
\label{literature}
The accurate forecasting of container logistics is pivotal for optimizing the operational efficiency and strategic planning of ports and shipping companies. Recent advancements in forecasting methodologies, particularly leveraging machine learning and data analysis techniques, have shown promising outcomes in tackling the complexities inherent in container logistics. This section provides an overview of the forecasting methods and models utilized in the container demand forecasting literature. In the area of predicting empty container availability, there is a focused but significant body of research dedicated to optimising repositioning strategies.\\

Diaz et al. \cite{diaz2011forecasting} underscored the significance of employing methodologies for precise forecasting of port empty container volume to achieve more cost-effective plans for empty container repositioning. Their study involved a comparative analysis of forecasting methodologies, including the Tioga Group, United Nations, and Winters methods, to predict the volume of empty containers at prominent US container ports, specifically the Port of Long Beach, Port of Los Angeles, and Port of Savannah. Liu Yuan \cite{liu2019machine} explored the relative performance of machine learning algorithms compared to conventional methodologies such as the Tioga Group, United Nations, and Winters methods in forecasting empty container volumes. Their research demonstrated the superior forecasting capabilities of machine learning algorithms based on datasets from the Los Angeles Port and Long Beach Port. Llopis et al. \cite{meseguer2021artificial} introduced an innovative approach for forecasting future 7-day stock using Artificial Neural Networks to predict the reception and withdrawal of empty containers in depots. The authors utilized various messages generated throughout the containers' shipment journey, along with temporal data associated with these events, to construct their predictive model.\\

Cai et al. \cite{cai2022data} developed a data-driven framework to tackle the Empty Container Repositioning (ECR) challenge by leveraging extensive datasets. This framework employed machine-learning algorithms to predict the supply and demand of empty containers and to identify critical parameters for incorporation into optimization models. Utilizing these parameters, the authors proposed two optimization models tailored for different ECR modes, with the aim of devising optimal ECR strategies. In a similar vein, Martius et al. \cite{martius2022forecasting} introduced innovative methodologies aimed at improving empty container relocation decisions and reducing unnecessary transportation costs. Their proposed approaches utilize artificial neural networks and mixture density networks to forecast the future weekly availability of empty containers globally. Specifically, their methodologies focus on predicting the future movements of individual containers, thereby enabling insights into container availability across numerous locations worldwide. Additionally, Sommer Benedikt \cite{sommer2023forecasting} presented a framework for forecasting empty container returns and deliveries. They emphasised the critical importance of accurate forecasting of demand for and return of empty containers at various spatial and spatial and temporal dimensions to facilitate informed repositioning decisions. These decisions rely heavily on accurate estimates of future empty container volumes with a forecasting up to 13 weeks in advance.

\section{Methodology}
\label{methodology}
The empty container availability forecasting system, a key component of the VBS backend modules, is designed to support carrier operations planning with hourly forecasts. It processes historical data on an hourly basis to meet the VBS's operational demands. This system predicts empty container availability over a five-business-day horizon, facilitating proactive transport planning and scheduling. Historical data are used to develop and validate these forecast models.

   \begin{figure}[h!]
      \centering
      \includegraphics[scale=0.5]{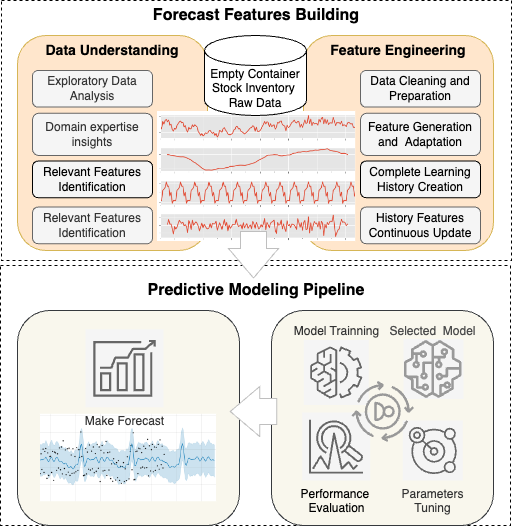}
      \caption{Proposed Approach for Forecasting Empty Container Availability.}
      \label{fig1}
     \vspace{-4mm}
   \end{figure}
These observations underpin the development of forecasting models that effectively capture the dynamic patterns of container availability at terminal depots. The forecast system, acknowledging the variety of container types and shipping companies, delivers customized forecasts to meet specific operational needs. Figure \ref{fig1} outlines the architecture of the forecasting approach, detailing component interactions and data flows that integrate historical data, models, and operational parameters to provide actionable insights for efficient empty container management. This study evaluates empty container forecasting using four models: the naive approach, ARIMA (AutoRegressive Integrated Moving Average) \cite{stevenson2007comparison}, Prophet \cite{taylor2018forecasting}, and LSTM (Long Short-Term Memory) \cite{greff2016lstm}. Each model is chosen for its relevance to time series forecasting in a VBS, with details on their mathematical foundations and implementation explained, providing a basis for further performance analysis.

\subsection{Problem Definition}
\label{problem}
The primary objective of this study is to enable transporters to proactively identify and leverage opportunities at the container terminal to enhance their operational efficiency during visits. This necessity has prompted the development of a forecasting approach specifically designed for this context. The main challenge addressed in this research is the accurate prediction of hourly container availability, considering the various types of empty containers at the terminal's depot. We aim to establish a forecasting framework that captures the complex dynamics of container availability, allowing carriers to effectively plan and optimize their operations in alignment with terminal schedules.
\subsection{Naive method}

The naive forecasting method stands as one of the simplest and most intuitive approaches utilized in time series analysis. It serves excellently as a baseline for more intricate models, owing to its straightforward applicability in scenarios where the data series does not exhibit significant trends or seasonal patterns.

The core principle of the naive method posits that the forecast for any future period is assumed to be identical to the last observed value. This concept can be encapsulated mathematically as follows:

\[
\hat{y}_{t+1} = y_t
\]

where:

\begin{itemize}
    \item $\hat{y}_{t+1}$ represents the forecast for the next period,
    \item $y_t$ denotes the actual observed value at time $t$.
\end{itemize}

\subsection{ARIMA}
An ARIMA model, short for Autoregressive Integrated Moving Average, is a univariate time series model aimed at capturing the dynamics of a single variable \cite{stevenson2007comparison}. In a such model, the time series is represented as a combination of its autoregressive (AR), Integration (I), and moving average (MA) components. The parameters \(p\), \(d\) and \(q\) respectively denote the orders of the AR component, integrated series, and MA components, reflecting the number of lagged observations considered for prediction, the degree of integration required to achieve stationarity, and the number of lagged forecast errors included in the model. The AR component operates under the premise that future observations can be estimated and forecasted based on the behavior of past and current values. Conversely, the MA component is designed to capture processes where the effects of previous environmental innovations persist and influence outcomes over a defined number of periods. In an ARIMA $p, d, q$ process, ensuring stationarity of the series used in the estimation process is essential. Accordingly, the series $y_t$  evolves according to the equation:

\[
y_t = \sum_{i=1}^p \beta_i y_{t-i} + \theta_0 + \sum_{j=1}^q \theta_j \varepsilon_{t-j}
\]

Here, $\beta_i$ refers to the AR coefficients,  $\theta_0$ represents a constant, $\theta_j$ denotes the MA coefficients, $\varepsilon$ denotes the error term, $q$ signifies the number of lagged terms of $\varepsilon$, and $p$ indicates the number of lagged terms of $y_t$.

\subsection{Prophet forecasting algorithm} 
The Prophet Forecasting Model introduced in the work proposed by Taylor et al. \cite{taylor2018forecasting} adopts a decomposable time series framework, comprising three primary model components: trend, seasonality, and holidays. These components are integrated into the following equation:
\[ y_t = g_t + s_t + h_t + \epsilon_t \]

Here \( y_t \) represents the forecasted value at time \( t \), \( g_t \) denotes the trend function, capturing non-periodic variations in the time series' value, while \( s_t \) represents periodic fluctuations such as weekly and yearly seasonality, and \( h_t \) accounts for holiday effects occurring on potentially irregular schedules over one or more days. The error term \( \epsilon_t \) represents any idiosyncratic changes which are not accommodated by the model; it is assumed to follow a parametric normal distribution. In Prophet, two trend models are considered: 
\begin{multicols}{2}
    \text{A saturating growth model:}
  \[g_t=\frac{C}{1+\exp (-k(t-m))} \]  
  \text{A piecewise linear model:}
  \[g_t=\left(k+\mathbf{a}(t)^{\top} \boldsymbol{\delta}\right) t+\left(m+\mathbf{a}_t^{\top} \boldsymbol{\gamma}\right) \]  
\end{multicols}
In the first trend model, \( C \)  represents the carrying capacity, \( k \) denotes the growth rate, and \( m \) serves as an offset parameter. In the second trend model,  \( k \) denotes the growth rate, \( \mathbf{a}_t \) refers to changepoint indicators where the growth rate can change, \( \boldsymbol{\delta} \) represents the rate adjustments, \( m \) is the offset parameter, and \( \boldsymbol{\gamma} \) signifies the correct adjustments at changepoints. The seasonality search leverages Fourier series to construct a flexible model of periodic effects with a considered regular period \(P\) that a time series could have, expressed as follows:
\[s_t=\sum_{n=1}^N\left(a_n \cos \left(\frac{2 \pi n t}{P}\right)+b_n \sin \left(\frac{2 \pi n t}{P}\right)\right)\]  

To incorporate holidays into the model, their effects are assumed to be independent. An indicator function is introduced to the model, indicating whether time \(t\) falls during holiday \(i\) and assigning each holiday a parameter \(\kappa_i\) representing the corresponding change in the forecast. This is illustrated in the following equations:
\[ Z(t)=\left[\mathbf{1}\left(t \in D_1\right), \ldots, \mathbf{1}\left(t \in D_L\right)\right]  \text{, } \quad h(t)=Z(t) \kappa \]

where, \( \kappa \sim \operatorname{Normal}\left(0, \nu^2\right) \) and \(D_i\) represents the set of past and future dates for each holiday \(i\).

\subsection{LSTM}
The traditional Recurrent Neural Network (RNN) models sequential data using network delay recursion, where the output at time 
\( t \) depends on both current and previous inputs. However, RNNs struggle with long-term context, crucial for tasks like time series forecasting \cite{pascanu2013difficulty}. The Long Short-Term Memory (LSTM) model, introduced by Hochreiter and Schmidhuber \cite{hochreiter1997long}, addresses this by managing long-term dependencies using memory cells and gate mechanisms. These memory cells function like conveyor belts, controlled by gates formed from sigmoid neural layers and pointwise multiplication, regulating information flow. At each time step \( t \), input \( X_t \) and the previous hidden state \( S_{t-1} \) are processed to compute the current hidden state \( S_t \), which is computed as follow :

\begin{enumerate}
    \item In LSTM, the initial step is determining which information to discard from the cell state. This choice is governed by the forget gate (\( f_t \)):
    \[ f_t = \sigma(X_tU^f + S_{t-1}W^f + b_f) \]
    
    \item Next, the process involves deciding which new information to incorporate into the cell state. This comprises two stages: firstly, the input gate (\( i_t \)) layer determines which values to update, and secondly, a hyperbolic tangent (tanh) layer generates a vector of new candidate values (\( \tilde{C}_t \)). These two stages can be outlined as follows:
    \[ i_t = \sigma(X_tU^i + S_{t-1}W^i + b_i) \]
    \[ \tilde{C}_t = \tanh(X_tU^c + S_{t-1}W^c + b_c) \]
    
    \item Following that, the previous cell state Ct-1Ct-1 is updated to yield the new cell state CtCt. This can be expressed as:
    \[ C_t = C_{t-1} \otimes f_t \oplus i_t \otimes \tilde{C}_t \]
    
    \item Lastly, the determination of the output is carried out. This output is derived from the cell state but in a filtered form. Here, the output gate (otot) determines which segments of the cell state will serve as the output. Subsequently, the cell state undergoes processing through a hyperbolic tangent (tanh) layer to constrain values between -1 and 1, followed by multiplication by the output gate. This process is outlined as follows:
\[ o_t = \sigma(X_tU^o + S_{t-1}W^o + b_o) \]
\[ S_t = o_t \otimes \tanh(C_t) \]

\end{enumerate}

From the preceding six equations, the LSTM architecture can be divised into three distinct groups of parameters:
\begin{enumerate}
    \item Input weights: \( U^f \), \( U^i \), \( U^o \), \( U^c \).
    \item Recurrent weights: \( W^f \), \( W^i \), \( W^o \), \( W^c \).
    \item Bias terms: \( b_f \), \( b_i \), \( b_o \), \( b_c \).
\end{enumerate}

LSTM offers a powerful tool for forecasting empty container availability in terminals. By analyzing historical data and capturing long-term dependencies, LSTM models enable accurate predictions, optimizing container allocation and terminal operations. Real-time data integration enhances forecasting precision, aiding proactive decision-making.

\subsection{Performance Metrics}
\label{performance_metrics}
Various techniques and metrics exist for assessing the efficacy of regression models, with the selection contingent upon the particular problem, dataset characteristics, and objectives \cite{botchkarev2018performance}. Among the frequently employed regression performance metrics are:

\begin{enumerate}
    \item {
        \textbf{Mean Absolute Error (MAE):} The MAE quantifies the mean absolute disparity between predicted and actual target values. Its calculation involves averaging the absolute differences between each predicted value and its corresponding actual value:
       \[MAE = \frac{1}{n} \sum_{i=1}^{n} |y_i - \hat{y}_i|\]
       Here,  \(n\) denotes the total number of data points, \(y_i\) signifies the actual target values, and \(\hat{y}_i\) represents the predicted values.
    }
    \item { \textbf{Mean Squared Error (MSE):} The MSE computes the mean squared deviation between predicted and actual values, assigning greater emphasis to larger errors. Its formulation is as follows:
    \[MSE = \frac{1}{n} \sum_{i=1}^{n} (y_i - \hat{y}_i)^2\] 
    }

    \item { \textbf{Root Mean Squared Error (RMSE):} The RMSE represents the square root of the MSE and is commonly favored when the objective is to present an error metric in the same units as the target variable. Its computation is expressed as:
       \[RMSE = \sqrt{MSE}\]
    }


\end{enumerate}

\subsection{Data}
\label{data}
The data set for this study includes real data from a terminal in a North American port covering the period from  01 January 2022 to 13 April 2024. It contains detailed information about the history of container traffic, such as the name of the shipping company, the ISO code of the container and the type of cargo, specifically distinguishing between empty and full containers. Additionally, each record includes timestamps indicating the container's entry and exit from the yard. These timestamps are essential for the accurate tracking of availability patterns. The initial dataset was refined to include only empty containers, with entries for full ones removed in order to focus on forecasting their availability. Each container's hourly presence was tracked and aggregated by ISO code, with the aim of analysing daily stock levels and trends in the yard. This approach helped to identify patterns that were crucial for accurate forecasting. Containers were also classified into three groups according to their ISO codes, with the intention of streamlining analysis and improving model clarity.

\begin{itemize}
    \item \textbf{Standard Containers}: Including dry containers (regular and high-cube) and specialized types such as Ammo Grade, predominantly used for standard cargo.
    \item \textbf{Special Containers}: Comprising specific cargo types that do not conform to standard sizes, such as Tankers, Flatracks, Open Tops, and platforms for oversized cargo.
    \item \textbf{Reefer Containers}: Consisting of refrigerated containers (both regular and high-cube), essential for transporting temperature-sensitive goods.
\end{itemize}

\begin{figure}[h]
    \centering
    \includegraphics[width=0.8\textwidth]{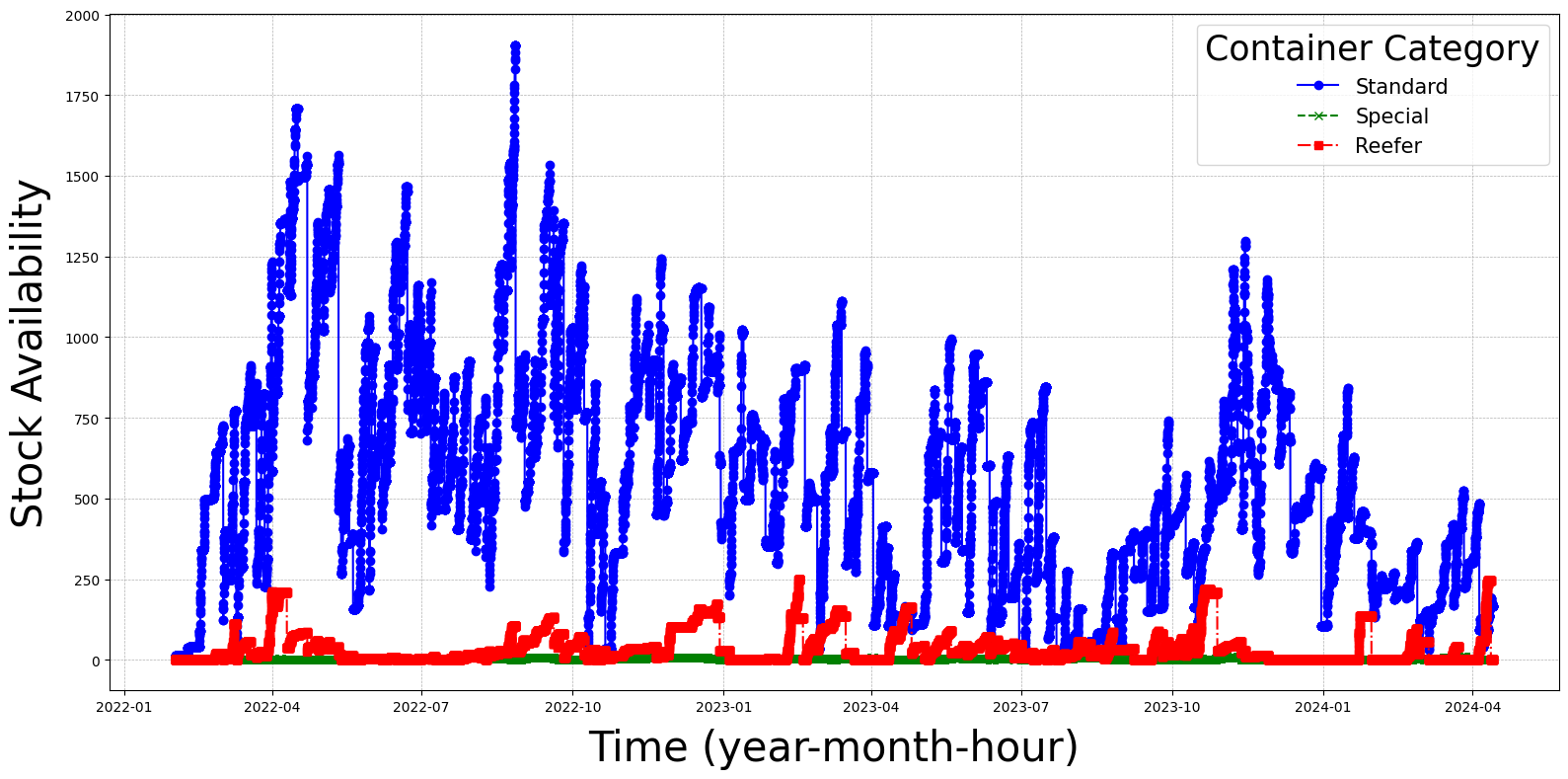}
    \caption{Trends in Empty Container Stock by Type Over Time.}
    \label{fig:container_stock_trends}
    \vspace{-4mm}
\end{figure}

Figure \ref{fig:container_stock_trends}. illustrates the trends in empty container stock by category from 1 January 2022 to 13 April 2024. It traces the hourly fluctuations in stock levels of Standard, Special, and Reefer containers, thereby providing a visual representation of their availability. Reefer and Special containers exhibit consistently low availability due to high demand and limited supply, which leads to frequent turnovers. Consequently, we will exclude Special and Reefer containers from our analysis and focus on Standard containers, which exhibit higher and more stable stock levels. This focus will enhance the precision and relevance of our forecasting models.

During the experimentation phase, data preparation entailed the organisation of the dataset into temporal folds for effective machine learning analysis. A time series-aware cross-validation approach has been adopted, which employs a sliding window method to divide the dataset into monthly segments, extending back five months from the most recent data point. The final seven days of each interval, comprising five business days, constituted the test set, while the earlier days formed the training set. This cross-validation approach is of vital importance for testing the forecasting models on unseen data and maintaining the sequence of real-world events. This allows for an assessment of the model's generalization across varying time conditions and the handling of seasonal changes.

\subsection{Forecasting Model Development and Parameter Tuning
}
\label{parameter_tunning}

To establish a rigorous foundation for our forecasting analysis, we employed a meticulous methodology that combined traditional time series analysis with advanced parameter optimization techniques for ARIMA, Prophet, and LSTM models. Each model was tailored to the specifics of our dataset, ensuring precision and relevance in forecasting.

\subsubsection{ARIMA Model Configuration}
The experimentation began with an assessment of the stationarity of the time series data, utilizing the Dickey-Fuller test. Initially, the test yielded a p-value of 0.071, indicating marginal stationarity. To enhance this, we applied a logarithmic transformation and differencing technique, reducing the p-value significantly to 9.21e-10, thus confirming improved stationarity. The identification of the ARIMA model parameters (p, d, q) was guided by the autocorrelation function (ACF) and the partial autocorrelation function (PACF) plots, from which we determined optimal values of p=2, d=1, and q=5.

\subsubsection{Prophet Model Configuration}
For the Prophet model, a grid search was conducted over a defined parameter space to find the best fitting model parameters. The search space included variations in changepoint prior scale, seasonality prior scale, and seasonality mode. The resulting optimal configuration is presented in the table below:

\begin{table}[h]
\centering
\begin{tabular}{|l|c|c|c|}
\hline
\textbf{Parameter} & \textbf{Changepoint Prior Scale} & \textbf{Seasonality Prior Scale} & \textbf{Seasonality Mode} \\
\hline
Values & 0.001, 0.01, 0.1, 0.5 & 0.01, 0.1, 1.0, 10.0 & Additive, Multiplicative \\
\hline
\end{tabular}
\caption{Prophet Model Grid Search Parameters.}
\label{table:prophet_params}
\end{table}

The optimal configuration was identified as a multiplicative seasonality mode, with a changepoint prior scale of 0.01 and a seasonality prior scale of 0.01. Accordingly, the model was set with yearly seasonality enabled and both weekly and daily seasonality disabled based on our observation of the dataset.

\subsubsection{LSTM Model Configuration}
A comprehensive grid search was also performed for the LSTM model, targeting key parameters that influence the network’s architecture and learning process. The parameters explored and the selected optimal settings are outlined in the following table:

\begin{table}[h]
\centering
\begin{tabular}{|l|c|c|c|}
\hline
\textbf{Parameter} & \textbf{Timesteps} & \textbf{Number of Epochs} & \textbf{Number of Layers} \\
\hline
Values & 60, 80, 100, 150, 200 & 50, 100, 120, 150, 200 & 2, 3, 4, 5, 8, 10 \\
\hline
\end{tabular}
\caption{LSTM Model Grid Search Parameters.}
\label{table:lstm_params}
\end{table}

The optimal LSTM parameters were determined to be 150 timesteps, 200 epochs, and 5 layers, which were implemented to maximize the model's predictive capability. Each model was carefully configured and trained using the aforementioned parameters, with each step in the model development process aligned to enhance forecasting accuracy and model robustness.

\section{Results and Discussion}
\label{results}

In this study, we evaluated the performance of four forecasting methods --Naive, ARIMA, Prophet, and LSTM-- to assess their efficacy in predicting empty container availability for a VBS application. The evaluation metrics employed included MSE, RMSE, and MAE, along with their corresponding standard deviations, providing a thorough evaluation of each model's predictive accuracy and consistency. The results of these evaluations are presented in Table \ref{tab:forecasting_performance}.

\begin{table}[ht]
\centering

\label{tab:forecasting_performance}
\begin{tabular}{@{}lcccccc@{}}
\toprule
\textbf{Method} & \textbf{Mean MSE} & \textbf{Std MSE} & \textbf{Mean RMSE} & \textbf{Std RMSE} & \textbf{Mean MAE} & \textbf{Std MAE} \\ \hline
Naive   & 49,813 & 63,620 & 189 & 118 & 167 & 117 \\ \hline
ARIMA   & 37,425 & 28,997 & 182 & 66 & 139 & 63 \\ \hline
Prophet & 81,539 & 123,369 & 221 & 181 & 198 & 186 \\ \hline
LSTM    & \textbf{18,255} & \textbf{10,396} & \textbf{129} & \textbf{40} & \textbf{87} & \textbf{33} \\ \bottomrule
\end{tabular}
\caption{Summary of Forecasting Performance Metrics by Method.}
\end{table}

Our findings indicated significant variations in the performance of the methods. The LSTM approach exhibited outstanding performance across all assessed metrics, recording the lowest Mean MSE of 18,255, Mean RMSE of 129, and Mean MAE of 87. Its standard deviations were also the lowest among the methods (Std MSE of 10,396, Std RMSE of 40, and Std MAE of 33), suggesting high consistency in its predictions. In stark contrast, the Prophet method displayed the highest error metrics among all models, with a Mean MSE of 81,539 and a Mean RMSE of 221, indicating its significantly lower predictive accuracy. Remarkably, it even underperformed the Naive method. This suggests that the Prophet method may not be well-suited for this particular application. In comparison, the ARIMA model exhibited better outcomes, with performance metrics indicating it outperformed the Naive model in reducing Mean MSE and MAE. The superior performance of the LSTM model can be ascribed to its ability to capture and remember long-term dependencies within the dataset, a vital characteristic for managing the dynamic and irregular patterns typical in container availability forecasts. On the other hand, the high variability in Prophet's performance, indicated by its substantial standard deviations, suggests a potential overfitting to noise rather than capturing the underlying trend effectively, which could undermine its reliability in practical applications.
Moreover, while the ARIMA model is traditionally favored for its ability to model a broad range of behaviors in time series data, it was outperformed by the LSTM. This outcome can be attributed to ARIMA's limitations in managing non-linear patterns, which are better handled by LSTM's deep learning capabilities. These results underscore the effectiveness of advanced machine learning models like LSTM in complex forecasting tasks, suggesting their significant utility in logistics and supply chain management, particularly for operational efficiency and cost management. Future work could explore combining the predictive strengths of traditional statistical models and advanced machine learning techniques in an ensemble approach to further enhance accuracy and reliability. Additionally, leveraging more detailed real-time data on container usage might refine these models' forecasting capabilities further.
Overall, our analysis recommends the adoption of advanced predictive analytics in forecasting empty container availability as a strategic enhancement for more robust and efficient operations within VBS and the broader logistics industry.

\section{Conclusion}
\label{conclusion}
This study has critically examined the performance of four different forecasting methods—Naive, ARIMA, Prophet, and LSTM—in predicting empty container availability at container terminal depots, employing a VBS for a future time horizon of five business days. The efficacy of these methods was evaluated using comprehensive metrics such as MSE, RMSE, and MAE, alongside their respective standard deviations. The results demonstrated that the LSTM model outperformed the other methods, exhibiting superior predictive accuracy and consistency in forecasting empty container availability. This highlights the suitability of advanced machine learning techniques over traditional methods for handling complex, dynamic patterns observed in container terminal operations. Conversely, the Prophet model, despite its general applicability in various forecasting scenarios, performed poorly compared to even the Naive method in this specific context. This underperformance may be attributed to its sensitivity to outliers and noise in the dataset, which are common in the operational data of container terminals. The findings of this research contribute to the improvement of operational efficiencies at container terminals through the optimization of empty container management. Furthermore, they highlight the importance of selecting appropriate forecasting techniques that align with the specific conditions and needs of the application. As global trade continues to expand and the complexity of supply chain networks increases, the ability of container terminals to effectively predict and manage empty container availability becomes even more crucial. Future studies could investigate the integration of ensemble methods that combine the strengths of traditional statistical models with machine learning techniques to further enhance forecasting accuracy. Additionally, the inclusion of real-time data and more granular details about container movements could provide further improvements in predictive performance. A future scope of this research will delve deeper into the granularity of predictions, specifically forecasting empty container availability per ISO type and shipowner, which is particularly relevant for enhancing trucking operations. By continuing to refine these forecasting models, stakeholders in the logistics and supply chain sector can better anticipate challenges and efficiently manage resources in a rapidly evolving global market.





\bibliography{bibliography.bib}




\clearpage

\end{document}